\documentclass[12pt]{article}
\usepackage{graphicx}
\usepackage{setspace}
\usepackage[margin=1in]{geometry}
\usepackage{longtable}
\usepackage{multirow}
\usepackage{amsmath} 
\usepackage{authblk}
\usepackage{cite}
\usepackage{float}
\usepackage{physics}
\usepackage{url}
\usepackage[round]{natbib}

\title{\textbf{Enhancing Trading Performance Through Sentiment Analysis  with Large Language Models: Evidence from the S\&P 500}}

\author[1]{Haojie Liu}
\author[2]{Zihan Lin}
\author[2]{Randall R. Rojas}

\affil[1]{Department of Statistics and Data Science }
\affil[2]{Department of Economics \break University of California, Los Angeles}

\begin{document}

\begin{singlespace}
\maketitle
\end{singlespace}

\begin{abstract}

This study integrates real-time sentiment analysis from financial news, GPT-2 \citep{radford2019language} and FinBERT \citep{araci2019finbert}, with technical indicators and time-series models like ARIMA and ETS to optimize S\&P 500 trading strategies. By merging sentiment data with momentum and trend-based metrics, including a benchmark buy-and-hold and sentiment-based approach, is evaluated through assets values and returns. Results show that combining sentiment-driven insights with traditional models improves trading performance, offering a more dynamic approach to stock trading that adapts to market changes in volatile environments.

\end{abstract}

\section{Introduction}

In the stock market, one of the most important factors to rely on is a strategy that consistently predicts market movements. Traditionally, such strategies have focused on technical indicators and time series models, which analyze historical price data and trends to forecast future stock movements. However, with the rise of Natural Language Processing (NLP), sentiment analysis has emerged as a valuable addition to financial models, offering an additional layer of input to enhance predictability. News articles and financial reports, whether positive or negative, often influence investor behavior, and sentiment analysis provides a way to quantify the sentiment embedded in these sources of information \citep{huang2018news}.

Sentiment analysis has brought about a significant breakthrough in financial forecasting. Among these time-series models are ARIMA, ETS, and Prophet, which have long served as key tools in successfully predicting other time-series data drawn from historical trends \citep{shumway2025tsa, hyndman2018forecasting}. However, these depictions struggle to explain the impact of real-time events and news \citep{xu2023emh}, which can often drastically change market dynamics. Fama's Efficient Market Hypothesis (EMH) contends that since stock prices incorporate all available information, investors cannot consistently earn blockbuster returns; however, differentiating between weak, semi-strong, and strong forms of market efficiency exposes how heuristics can help us understand why market reactions to real-time news remain partially incomplete in many instances relative to the assumptions of traditional models.

Previous studies extend this by incorporating sentiment data from different channels such as Twitter [X] and news articles \citep{gu2020twitter, kirtac2024llm} to improve trading strategies using market sentiment in addition to historical price data. A study by \citep{gu2020twitter} demonstrated how emotional tweets on X can be directly linked to stock market indicators, providing evidence that following public sentiment helps in real-time tracking of such events. In a similar fashion, \citep{kirtac2024llm} showed that media pessimism, measured by large-language-model sentiment scoring, is predictive of downward market pressure, suggesting that the instantaneousness of media sentiment can affect investor behavior. This combination of sentiment analysis and technical forecasting methods could present a richer and more volatile context for market behavior than traditional models alone.

Sentiment analysis is a method for determining sentiment from written sources, either positive or negative. When used in financial news, this could offer exposure to market sentiment and consequently to stock prices. In a related finding, \citep{greyling2025twitter} demonstrated that collective mood states inferred from X data—categorized as positive or negative emotions—significantly enhanced the accuracy of predicting whether the Dow Jones Industrial Average would rise or fall the next day. As a result, their study identified the public mood dimensions which are calculated based on public tools like OpinionFinder (OF) \citep{finvader2023tool} and Google-Profile of Mood States \citep{poms2021short}, which lead to better prediction rates that finally mean, that real-time sentiment could affect stock market behavior.

Technical analysis, on the other hand, uses indicators or calculations based on price and volume (momentum as well) that measure volume and momentum to identify trends and turning points in stock prices. For instance: \citep{moreno2020sentiecon} and \citep{du2024fsa} have stressed the true significance of text categorization related to financial literature; elucidating the plain fact that common financial terms are often miss-categorized by previous dictionaries utilized for textual analysis. This underlines the importance of financial sentiment of fine-tuned tools to enable the fine tuning required to convey negative or positive word usage that might impact trading decisions. Integration of these technical indicators with sentiment analysis can help in filtering market trends, and hence, may improve trading signals, and ultimately prediction.

Similarly, \citep{critien2022bitcoin} studied the cross-sectional relationship between sentiment and financial asset pricing using sentiment on X along with time-series analysis for forecasting Bitcoin prices. They found a strong link between public sentiment (in X) and Bitcoin short-run price momentum, showing that using social media indicators with time-honored economic predictors and technical variables, this concept can have better utility in price forecast models. Their work provides some useful learnings that make one think about whether sentiment analysis could bolster traditional time-series models in terms of predictive power — particularly when applied to real-time data. Thus, this study seeks to expand the aforementioned studies to the S\&P 500 index by employing sentiment with technical data and time-series forecasting models to improve the accuracy of stock return predictions and trading performance.

To get a more accurate perspective on the market sentiment, it is possible to combine the sentiment score with other technical indicators. For analyzing financial text, domain-specific models like FinBERT have been created to capture some of the nuances in language that are inherently part of finance. \citep{yang2019finbert} and \citep{liu2020finbert} proved that FinBERT, a model pre-trained on extensive financial text corpora, provides better predictions about the trends in stock prices than general natural language processing models. This goes on to show that combining sentiment analysis from financial news and technical indicators might be an even better way to devise trading strategies and glean insights into the market.

Our research is aimed at assessing the effect of sentiment on stock trading strategies, and time-series forecasting models. In this work, in particular, address the question if sentiment information from news articles can be used to obtain an improvement in stock return prediction and to achieve better overall trading performance on the S\&P 500 market. This is consistent with prior work that has shown that applying text-based classification methods, like support vector machines, to news articles can improve the abnormal returns to predict hits concerning magnitude but not necessarily direction \citep{fazlija2022news}. Their work offers insights into combining textual data with market indicators to improve financial predictions.

We mainly develop the sentiment analysis of financial news to improve trading strategies and time-series models for stock returns. Through the capture of ongoing market sentiment and coupling that with a historical perspective of price action, this hybrid predictive method performs better than classic methods which only use technical indicators or price-only predictions. And other research does bolster this theory. For instance, \citep{kirtac2024llm} observed that negative media sentiment—notably an increased pessimism—can predict future downward pressure on stock prices which tend to revert to fundamentals. The result indicates that sentiments are not just noises, but they can drive trading volume and price movements.

Additionally,  \citep{fazlija2022news} found out that sentiment analysis on news data improves stock price return prediction compared to the traditional bag-of-words model. Constructing a dictionary based on a financial domain-specific lexicon, they formed sentiment spaces in various market levels (stock, sector, index), this performed better than purely statistical methods. Together, these studies imply that incorporating sentiment with time-series models can improve the accuracy and stability of stock return forecasting, providing traders with a more comprehensive picture of market behaviors.

We merge news sentiment data from a variety of sources and languages, then processed using state-of-the-art NLP building blocks (such as GPT or FinBERT) with the traditional financial data from the S\&P 500. These sentiment analysis outputs are then subsumed with technical indicators like MACD, SAR, Volume Weighted MACD, DUAL MACD; as well as time-series models like ARIMA, Prophet, and ETS. A significant amount of moving averages are based on the last adjusted closing price and we use these closing prices for our calculations. To highlight an example; MACD — a common momentum indicator in trading that heavily uses the exponential moving average (EMA) over its counterpart — will be more prone to consider recent price changes than potentially justifiable alongside simple moving average (SMA). In \citep{wang2018predicting}, the predictive accuracy of MACD (Moving Average Convergence Divergence) is significantly enhanced by integrating it with HVIX (Historical Volatility Index), a metric that quantifies the degree of price variability or uncertainty in a stock over a historical period. By leveraging HVIX, the strategy can better adapt to market conditions, as the index provides critical insights into periods of high or low volatility. This integration enables more precise identification of stock trends and optimizes trading decisions. As a result, this combined approach outperforms traditional MACD-based strategies, delivering improved returns across different time horizons by 33.33\% and 12\%, respectively.

In response to the criticism of MACD, \citep{chio2022macd} applied MACD to a wide range of stocks across markets, demonstrating its continued effectiveness. In some of the stock markets, such as the Milan Comit General and S\&P/TSX Composite Index, their analysis consistently created statistically significant abnormal returns from both MACD and RSI. These technical indicators also demonstrate the opportunity for combining them with sentiment analysis to improve trading strategies.

We develop and test a comprehensive set of trading strategies, including a benchmark buy-and-hold strategy and various sentiment-based strategies. Performance is evaluated based on the comparison of the last assets value cumulative return between strategies, aiming to determine whether combining sentiment data with technical indicators leads to higher returns and improved trading performance.

\section{Data}

In this study, we utilize two key datasets: financial news articles collected from sources such as the Wall Street Journal (WSJ), Barron, Benzinga, MarketWatch, and Dow Jones, along with daily stock prices for the S\&P 500 index (symbol: \^{}GSPC
). Both the news and stock data are collected at a daily frequency, providing a synchronized view of market sentiment and performance. These datasets form the basis for our sentiment analysis and trading strategy development. The financial news sources are crucial for understanding market sentiment, offering a broad and comprehensive perspective on investor attitudes.

Historically, sentiment derived from mainstream financial news outlets has had a measurable impact on stock prices. For example, \citet{tetlock2007giving} found that sentiment from news sources like the Wall Street Journal could predict stock price movements, highlighting the importance of these sources in shaping market perceptions. However, as \citet{loughran2011when} pointed out, misclassifications in financial texts can lead to inaccurate market signals, emphasizing the need for precise sentiment analysis to avoid incorrect conclusions.

Other studies, such as \citet{schumaker2009textual}, demonstrate that breaking news from sources like Reuters and Bloomberg can result in immediate market movements due to their in-depth coverage of critical events. Furthermore, \citet{engelberg2011causal} noted that both formal media outlets like Bloomberg and online message boards have a marginal influence on stock prices, with more formal sources having a greater immediate impact.

In this study, financial news from a variety of sources is combined with historical stock data from the S\&P 500 index, which includes key variables such as the open, high, low, close, and adjusted closing prices, as well as trading volume. The adjusted closing prices, in particular, are used to calculate daily returns, which are then aligned with sentiment derived from the news articles to inform our trading strategy. Below, we describe the datasets and the transformations applied to prepare the data for modeling.

\subsection{Stock Data: S\&P 500 Prices}

The stock data contains historical price information for the S\&P 500 index. We downloaded this data from Yahoo Finance, capturing daily prices from August 08, 2019 to August 07, 2024. The dataset includes the open, high, low, close, adjusted closing prices, and trading volume. Our primary focus is on the \textit{adjusted closing price}, which reflects the stock’s value after accounting for corporate actions such as dividends and stock splits.

Once the data was loaded, we transformed it by calculating \textit{daily returns} to quantify price changes. Specifically, for each day, we compute the percentage change in the adjusted closing price relative to the previous day. This return value gives us a normalized measure of the market’s movement. The formula for daily returns is as follows:

\begin{equation}
\text{Return}_t = \frac{\text{Adjusted Closing Price}_t - \text{Adjusted Closing Price}_{t-1}}{\text{Adjusted Closing Price}_{t-1}}
\end{equation}

Next, we categorized these daily returns into three classes, assigning each a sentiment label as follows:

\begin{itemize}
    \item \textbf{Positive (1)}: If the return is greater than 1\% (\textgreater 1\%).
    \item \textbf{Neutral (0)}: If the return is between -1\% and 1\%, inclusive (-1\% \(\leq\) return \(\leq\) 1\%).
    \item \textbf{Negative (-1)}: If the return is less than -1\% (\textless -1\%).
\end{itemize}

These labels form the stock market sentiment that will be used later to compare with sentiment derived from the news articles.

\subsection{Lag Structure and Weekend Handling} 

Articles time-stamped between 16{:}00 and 23{:}59~ET are linked to the \emph{next} trading day to respect the market close.  
All stories published on Saturday or Sunday are aggregated and paired with Monday’s return, following \citep{greyling2025twitter}.  To test robustness we estimate models with lags $k\!=\!0,1,2$; the $k=1$ specification (news today $\rightarrow$ return tomorrow) delivers the best out-of-sample accuracy.

\subsection{Financial News Data}

We gathered a collection of financial news articles from the Wall Street Journal (WSJ), Barron, Benzinga, MarketWatch, and Dow Jones, focusing on content that could influence stock market behavior. The dataset consists of three main columns:

\begin{itemize}
    \item \textbf{Date}: The publication date of the article.
    \item \textbf{Title}: The title of the article.
    \item \textbf{Text}: The full body of the article.
\end{itemize}

Since these news data are unstructured, we applied several preprocessing steps to clean and format it for analysis:

\begin{itemize}
    \item \textbf{Date Extraction}: Some of the article texts included the date embedded within the content itself. Using regular expressions, we extracted this date and standardized it across all rows.
    \item \textbf{Removing Non-News Data}: Certain rows in the dataset contained irrelevant content, such as articles mentioning ``Amundi S\&P 500,'' which was not pertinent to our analysis. These rows were identified using keyword matching and removed from the dataset to ensure the content remained relevant to financial news impacting the stock market.
    \item \textbf{Text Cleaning}: The article texts often contained unnecessary copyright information and footnotes, such as disclaimers or unrelated post-article content. We applied a cleaning function to remove this extraneous information, including terms like ``Copyright,'' ``Photo by,'' and various footer disclaimers. This step ensures that the text is solely focused on the main news content, making it more reliable for sentiment analysis.
\end{itemize}

\subsection{Data Integration and Sentiment Labeling}

Once the stock returns and news data were prepared, we merged these two datasets by the \textit{Date} column. This allowed us to align daily stock market movements with the corresponding news sentiment for the same day.

After merging, the stock price data was converted into categorical sentiment labels (positive, neutral, negative) as described earlier. We then appended this stock sentiment to each news article for that day. The final dataset includes both the news content and the associated stock market sentiment, enabling us to analyze the correlation between financial news and market performance.

For sentiment analysis, we used large language models like GPT-2 and FinBERT to assign sentiment scores to the cleaned news articles. The combined sentiment data (news-derived and stock price-derived) forms the basis of our trading strategy, where the goal is to leverage sentiment to make more informed trading decisions.

\section{Indicators}

In this section, we conducted a comprehensive trading simulation to evaluate the effectiveness of our integrated sentiment analysis and technical indicators in real-world trading scenarios. This simulation aimed to assess how well the combination of sentiment scores from GPT-2 and FinBERT, along with technical and time-series models, could predict stock price movements and generate profitable trading signals. This section details the methodology of the trading simulation, including the dynamic trading strategy employed, the integration of sentiment and technical indicators, and the benchmarking against a standard buy-and-hold strategy to provide a clear comparison of performance.

\subsection{Sentiment Indicators}

The sentiment analysis involved two models: GPT-2 and FinBERT, both of which were modified by adding a dropout layer to reduce the risk of overfitting. These models were used to analyze the textual content of financial news articles and produce sentiment scores, which were then integrated with stock price data for further analysis.

\subsubsection{GPT-2}

As a generative language model, GPT-2 has become popular for its ability to handle huge amounts of textual information and produce meaningful text from an input prompt. This research focuses on fine-tuning GPT-2 for the financial domain, which means it is trained on news data from daily articles. One key feature is its ability to generate sentiment scores for each article, providing valuable insights into market sentiment.
It can discern whether financial news items have an optimistic, neutral, or pessimistic sentiment and is thus key here to give trends of investor sentiment. Mimicked prices can be correlated with the movements in the stock market, which provides profitable signals for predicting stock prices. The use of GPT-2 aids the research through its ability to comprehend sophisticated financial vocabulary as well as the development of a sentiment interpreter, allowing for real-time market-tailored sentiment indicators that can change stock price directions.

\subsubsection{FinBERT}

FinBERT is primarily pre-trained on a financial corpus and hence more suitable for sentiment analysis in the finance domain. Instead of general-purpose language models, FinBERT is fine-tuned to be particularly effective at understanding complicated financial lexicon, and at context-specific sentiments like risk and uncertainty that we often find in financial news. In this study, FinBERT is used for the same news articles as GPT-2 and then we compare the output difference of words from these models and how much they predict the stock price movement.

This is because FinBERT is pre-trained on financial text and fine-tuned for sentiment analysis hence guaranteeing high precision especially since it is adapted to financial market conditions. This specialization is necessary so that we can have more accurate sentiment scores when we process the news, crucial for the generation of trading signals on which actions are made. By integrating FinBERT into our analysis, we benefit from a model that understands the nuanced language used in financial media, making our sentiment-driven trading strategies more robust. With the accurate, in-context sentiment scores derived from FinBERT, this can be considered a three-step solution and will help improve stock price prediction and trading strategies which is especially powerful when paired with time-series models.

\subsection{Technical Indicators}

In the realm of financial trading, technical indicators play a crucial role in analyzing market trends, identifying potential entry and exit points, and enhancing the overall decision-making process for traders. This section delves into several key momentum and trend-following indicators that are widely utilized in trading strategies. Specifically, we explore the Moving Average Convergence Divergence (MACD), Stop and Reverse (SAR), Volume-Weighted MACD (VW MACD), and Dual MACD indicators. Each of these tools offers unique insights into market dynamics, allowing traders to better interpret price movements and volatility. By understanding and effectively applying these indicators, traders can improve the accuracy of their predictions and optimize their trading performance.

\subsubsection{MACD (Moving Average Convergence Divergence)}

MACD (Moving Average Convergence \& Divergence) is a momentum based indicator that tracks the difference between the 12-days and 26-days exponential moving averages of a stock's price. The MACD line is derived by finding the disparity between these 2 averages and a signal line (a 9-days EMA) is then calculated which acts as confirmation for buying or selling signals. Buy Signal: When the MACD line crosses above the signal line, indicates a move to the upside Sell Signal: 

\setlength{\abovedisplayskip}{0pt}
\setlength{\belowdisplayskip}{5pt}

\begin{gather}
\text{MACD Line} = \text{EMA}_{12} - \text{EMA}_{26} \\
\vspace{-4mm} 
\text{Signal Line} = \text{EMA}_{9}(\text{MACD Line}) \\
\vspace{-4mm} 
\text{MACD Histogram} = \text{MACD Line} - \text{Signal Line}
\vspace{-4mm}
\end{gather}

When it crosses below its signal line, signals indicate that it is probably time to sell and expect down. As the indicator can be used for detecting the trend reversals it is more helpful to have short-term or medium-term trade strategies \citep{appel2005technical}.

\subsubsection{SAR (Stop and Reverse)}

SAR (Stop and Reverse), is a trend-following indicator which means if the current trend is bullish this can be called a bullish signal. 

\setlength{\abovedisplayskip}{0pt}
\setlength{\belowdisplayskip}{5pt}

\begin{gather}
\text{SAR}_{\text{current}} = \text{SAR}_{\text{previous}} + \alpha (\text{EP} - \text{SAR}_{\text{previous}}) \\
\vspace{-4mm} 
\text{SAR}_{\text{current}} = \text{SAR}_{\text{previous}} - \alpha (\text{SAR}_{\text{previous}} - \text{EP})
\end{gather}

EP stands for the Extreme Point, while the adjustment factor $\alpha$ increases over time as trends persist. It plots a row of dots: either at the low - under the price or above - when an upward trend is detected. For a long trade, SAR dots are placed above the price and would signal a potential reversal if the price breaks through the SAR dots, generating a sell signal. On the flip side, when SAR dots form above the price in a downtrend, they break out of that formation to trigger a buy signal. This indicator works great if you use it to participate in profits during the trend and prevent a nasty reversal \citep{kaufman2013trading}.

\subsubsection{VW MACD (Volume-Weighted MACD)}

Volume-Weighted MACD is a variation of the MACD indicator that accounts for volume. In very liquid markets you will often see large volume increases with price movements and this is something that VW MACD can act on to help improve the reliability of trading signals. The volume-weighted detail assures that peaks of high activity are offered more importance in momentum calculation so that purchasers can determine whether to act based on price movements on a volume foundation or if they will die out fast. This makes it an important tool for traders who are concerned with liquidity, and this indicator gives you even cleaner buy and sell signs based on reversals in such situations where volume comes into play as a crucial factor of market dynamics. In this study, when the price moves above the VW MACD level, it indicates a potential upward trend, suggesting a buy signal, otherwise sell signal.

\subsubsection{Dual MACD}

The Dual MACD is the same as VW MACD but instead uses two different MACD signals, usually a short-term MACD (e.g. 12-26-9 days) and then a longer-term MCD (e.g. 19-39-9 days). The Dual MACD attempts to solve this conundrum by confirming an indicator (the current time frame) with another one, providing a better filter for false signals and more confirmation of the direction of the trend. A signal is thought to be stronger when each of the quick-term and lengthy-term traces show an identical momentum exchange. This strategy will prevent us from getting into trades based on only short-term volatility trapping us in trends that are not real and it gives a lot more strength to our trend detection mechanism. It is particularly useful in markets that reverse regularly, preventing traders from entering or exiting too early.

\subsection{Time-Series Models}

Time-series forecasting models are powerful tools for analyzing and predicting stock price movements. In this section, we explore three widely used models: ARIMA, Prophet, and ETS. Each model offers unique strengths in capturing patterns, trends, and seasonality in stock price data, which, when combined with sentiment analysis, provide a comprehensive framework for generating informed trading signals. These models are applied to historical stock data to uncover insights into market behavior and enhance the effectiveness of our trading strategies.

\subsubsection{ARIMA}

One of the most commonly used methods for time-series forecasting is ARIMA Model (Auto\-Regressive Integrated Moving Average) and it works well with financial data such as stock prices \citep{box1970time}. When investigating this, it is used on historical stock price data to predict how the stock will move in the future, which in turn can aid in creating a strategy that trades upon these anticipated trends.

\setlength{\abovedisplayskip}{0pt}
\setlength{\belowdisplayskip}{5pt}

\begin{gather}
y_t = \phi_1 y_{t-1} + \phi_2 y_{t-2} + \dots + \phi_p y_{t-p} + \epsilon_t \quad (AR) \\
\vspace{-4mm} 
y'_t = y_t - y_{t-1} \quad (I) \\
\vspace{-4mm} 
y_t =  \theta_1 \epsilon_{t-1} + \theta_2 \epsilon_{t-2} + \dots + \theta_q \epsilon_{t-q} + \epsilon_t \quad (MA) \\
\vspace{-4mm} 
y'_t = \phi_1 y'_{t-1} + \dots + \phi_p y'_{t-p} + \theta_1 \epsilon_{t-1} + \dots + \theta_q \epsilon_{t-q} + \epsilon_t
\end{gather}

ARIMA is good for stock market data because it can handle the modeling non-stationary series and adapts to the trends and randomness in the data. ARIMA also helps our research by providing a model that allows us to make short-term price predictions with a reasonable amount of confidence. The addition of these predictions along with the sentiment analysis facilitates a much richer predictive approach to stock movement forecasting. Since we can optimize ARIMA’s parameters (p, d, q) (P, D, Q) on our data directly, this allows us to ensure that the model works for forecasts that align with our trading strategy.

\subsubsection{Prophet}

Prophet is a time-series forecasting model developed by Facebook \citep{taylor2018forecasting}, useful for dealing with datasets that exhibit seasonal trends and have non-regular collection rates. The real power of Prophet is in its flexibility and simplicity, most specifically its ability to model trends and seasonal component estimation separately. In our research, Prophet is used for predicting stock prices by decomposing the data into three components: Growth, Seasonal, and Holidays or Events.

\setlength{\abovedisplayskip}{0pt}
\setlength{\belowdisplayskip}{5pt}

\begin{gather}
y(t) = g(t) + s(t) + h(t) + \epsilon_t \\
\vspace{-2mm} 
g(t) = (k + a(t)^T \delta) \cdot t + (m + a(t)^T \gamma) \quad \text{(Piecewise Linear Trend)} \\
\vspace{-2mm} 
g(t) = \frac{C}{1 + \exp(-k(t - m))} \quad \text{(Logistic Growth Trend)} \\
\vspace{-2mm} 
s(t) = \sum_{n=1}^{N} \left[ a_n \cos\left( \frac{2 \pi n t}{P} \right) + b_n \sin\left( \frac{2 \pi n t}{P} \right) \right] \quad \text{(Seasonality Component)} \\
\vspace{-2mm} 
y(t) = g(t) + \sum s(t) + \sum h(t) + \epsilon_t \quad \text{(Full Prophet Model)}
\end{gather}

In markets, this can be an important element (think of a quarterly earnings announcement or a macroeconomic report) so the possibility for Prophet to model these effects means improved forecasts. Its simplicity and insensitivity to missing or outlier data also makes it very useful for long-term, seasonal trend prediction. Integrating Prophet into our research provides a model capable of handling the complexities of stock movements over time, complementing and adding depth to the periodic short-term fluctuations captured by models like ARIMA. In some regards, the Prophet outputs are combined with sentiment scores makes the indicator a more legitimate trading signals (it now accounted for both historical trends and real-time market sentiment.

\subsubsection{ETS}

Another advance time-series forecasting model is ETS (Error, Trend, Seasonality) \citep{hyndman2018forecasting}. If the data have strong cyclical patterns, ETS models are also great options as they allow you to capture the fundamental structural behavior of changes in stock prices. It is useful when we have a time-series, utilizing ETS to forecast a company's stock price. This approach captures the underlying market trend, any frequent seasonal variations, and accounts for randomness.

\setlength{\abovedisplayskip}{0pt}
\setlength{\belowdisplayskip}{5pt}

\begin{gather}
y_t = (T_t \cdot S_t) \cdot \epsilon_t \\
\vspace{-2mm} 
\epsilon_t = y_t - \hat{y}_t \quad \text{(Additive Error)} \\
\vspace{-2mm} 
\epsilon_t = \frac{y_t}{\hat{y}_t} \quad \text{(Multiplicative Error)} \\
\vspace{-2mm} 
T_t = T_{t-1} + b_{t-1} \quad \text{(Additive Trend)} \\
\vspace{-2mm} 
T_t = T_{t-1} \cdot b_{t-1} \quad \text{(Multiplicative Trend)} \\
\vspace{-2mm} 
S_t = S_{t - m} + S_{t - (m+1)} \quad \text{(Additive Seasonality)} \\
\vspace{-2mm} 
S_t = S_{t - m} \cdot S_{t - (m+1)} \quad \text{(Multiplicative Seasonality)}
\end{gather}

ETS is beneficial when we want to capture cyclical patterns in stock prices, which often arise due to economic cycles and shifts in market sentiment. This enables ETS to distinctly separate these components, providing unique insights into overall market dynamics that may enhance our trading strategy. By combining ETS predictions with sentiment scores from GPT-2 and FinBERT, we can create more sophisticated trading signals that leverage both endogenous stock price movements and exogenous market influences reflected in news sentiment. This integration helps make our simulations more informed and profitable overall.

\subsection{Analysis}

The implementation involved two programming languages: Python for large language models (LLMs) and R for trading strategies and time-series forecasting.

\subsubsection{LLM Setup}

The sentiment analysis was carried out in Python by using GPT-2 and FinBERT, leveraging major packages like Transformers, PyTorch, and Hugging Face's tokenizers. The transformers library allowed us to access those pre-trained models while PyTorch was how we trained and ran inferences on them. The text was preprocessed via tokenizers. Both GPT-2 and FinBERT were modified with a dropout layer to prevent overfitting: the dropout rate was set to 0.5 for GPT-2 and 0.8 for FinBERT. They fine-tuned these models for financial sentiment analysis and integrated the resulting sentiment scores with stock price data.

To ensure consistency between the sentiment analysis and stock price data, we aligned both on a daily frequency. Since there were three types of sentiment indicators—positive, neutral, and negative(+1, 0 and -1)—and multiple sentiment indicators could be generated for a single day, we applied a voting mechanism. The overall daily sentiment was determined by the category with the highest occurrence.

\subsubsection{Trading Strategy and Time-series Model Setup}

The trading strategy and time-series models were implemented in R. Technical indicators such as MACD, SAR, VW MACD, and Dual MACD were calculated using the \texttt{quantmod} and \texttt{TTR} packages. For time-series forecasting, models such as ARIMA, Prophet, and ETS were used, leveraging the forecast and prophet packages. The models were applied to daily historical stock prices.

\subsection{Trading Simulation}

A trading simulation was then carried out to assess how effective the sentiment analysis models were, based on technical indicators and time-series forecasting. It creates trading signals from the overall market sentiment and daily stock returns and utilize these for simulating trades to gauge real profitability.
Historical stock price data was used for this method as input to several technical indicators to help make trading decisions. These signals are selected to recognize market trends, changes in momentum and trend reversals — one of them being ideal for providing trading signals with more insight and precision.

The strategy is designed to adjust the assets dynamically based on predicted sentiment signals for each trading day. The analytical formulation of this strategy can be summarized as follows:

Initially, the asset starts with a capital of \$10,000 and holds no shares where $C_0 = 10000$. Also, representation of all other variables, $\mathbf{Sen:}$ Sentiment Indicatiors, $\mathbf{Tec:}$ Technical Indicators, $\mathbf{P:}$ current stock price, $\mathbf{S:}$ Share of Stocks, $\mathbf{C:}$ Current budget, and $\mathcal{R}$ Return. The simulation begins by creating an indicator set $\mathcal{I}$, which integrates both sentiment ($\mathbf{Sen}$) and technical ($\mathbf{Tec}$) indicators. The goal is to derive a combined signal $\mathcal{I}_t$, guiding trading decisions daily. For simplicity, this simulation assumes zero transaction costs and slippage. While this may slightly overestimate real-world returns, it allows for a cleaner comparison between the sentiment-based strategy and the buy-and-hold benchmark.

This signal is calculated as follows:
$$\mathcal{I}_t = \frac{\Sigma \mathcal{I}_{it}}{\abs{\Sigma \mathcal{I}_{it}}}$$

Here, the numerator represents the cumulative signal strength from all considered indicators on day $t$, while the denominator normalizes the result to ensure the signal falls within the range of $[-1,1]$. A positive $\mathcal{I}_t$ signals a buying opportunity, whereas a negative $\mathcal{I}_t$ suggests selling. When $\mathcal{I}_t = 0$, the strategy advises holding the current position.
Two critical asset variables, $\mathbf{C_{t+1}}$ and $\mathbf{S_{t+1}}$, are updated daily based on $\mathcal{I}_t$:
The capital $\mathbf{C_{t+1}}$ is updated depending on whether the sentiment suggests selling shares ($\mathcal{I}_t < 0$) or holding cash ($\mathcal{I}_t \geq 0$), assuming a sufficient budget exists:

$$
\mathbf{C_{t+1}} = 
\begin{cases}
    P_t * S_t & \text{if }  \mathcal{I}_{t} < 0\\
    C_t & \text{if }  \mathcal{I}_{t} \geq 0\\
\end{cases}
\text{when } \mathbf{C_{t}} > 0
$$

Similarly, share holdings $\mathbf{S_{t+1}}$ are adjusted to reflect either buying new stocks ($\mathcal{I}_t > 0$) or retaining the current number of shares ($\mathcal{I}_t \leq 0$), provided shares are available:

$$
\mathbf{S_{t+1}} = 
\begin{cases}
    \frac{C_t}{P_t} & \text{if }  \mathcal{I}_{t} > 0\\
    S_t & \text{if }  \mathcal{I}_{t} \leq 0\\
\end{cases}
\text{when } \mathbf{S_{t}} > 0
$$

Finally, the return $\mathcal{R}_t$ is computed as the ratio of the current capital $\mathbf{C_t}$ minus the initial capital $\mathbf{C_0}$ to the initial capital, serving as a measure of performance over time:

$$\mathcal{R}_t = \frac{\mathbf{C_{t}} - \mathbf{C_{0}}}{\mathbf{C_{0}}}$$

To evaluate the effectiveness of this dynamic strategy, a buy-and-hold strategy was employed as a baseline method for evaluating the performance of the sentiment-based and technical analysis-driven strategies. This strategy involved purchasing the S\&P 500 daily price at the start of the test period (May 10, 2024) and holding the position until the end of the test period (August 07, 2024), without making any intermediate trades. The buy-and-hold strategy served as a benchmark, enabling a straightforward comparison between this simple method and the more complex strategies developed using sentiment analysis, technical indicators, and time-series forecasting models.

\section{Results}

\subsection{Sentiment Classification}

Both a GPT-2 and FinBERT models were implemented to conduct sentiment classification using financial news articles from multiple sources. Results were grouped by day to represent a more accurate depiction of daily sentiment. The aggregated sentiment for each day was calculated using the mode of the predicted sentiments across all articles for that particular date. This was chosen as a relatively conservative parameter setting for picking how we reflect minimum sentiment throughout the day, but this mode is also to some extent representative of which sentiment is most typically predicted that day, making the day-by-day sentiment score more reliable. We also checked our first record sentiment for a reference in each group. This way it could represent the market sentiment of each day for trading strategies effectively.

The next step was to make sentiment results aggregative and incorporate them into trading strategies to study the influence of market sentiment on stock prices. This approach successfully captured the prevailing mood, as reported in a variety of sources, that was then applied to the subsequent trading analyses.

In Table 1, the accuracy comparison for sentiment classification across different news sources highlights the best-performing results for both models. FinBERT achieved its highest accuracy of 75.56\% on Benzinga, while GPT-2 performed best on WSJ with an accuracy of 65.48\%. These results emphasize the importance of selecting the most suitable model for specific data sources to capture sentiment effectively.

\subsection{Time Series Classification}

We evaluated the performance of various time series forecasting models in predicting financial trends. Specifically, we utilized the Auto-Regressive Integrated Moving Average (ARIMA), Prophet, and Exponential Smoothing State Space Model (ETS) to forecast daily stock returns.

In Table 2, Prophet and ETS models achieved the highest accuracy, both at 59.65\%, significantly outperforming ARIMA, which achieved an accuracy of 29.82\%. The superior performance of Prophet and ETS suggests that they are better suited for capturing the underlying patterns and seasonality in the financial time series data. These results indicate that time series models can be valuable tools in forecasting stock price movements, potentially enhancing trading strategies when combined with other indicators.

\subsection{Trading Indicator Classification}

In addition to time series models, we analyzed traditional technical indicators for their effectiveness in predicting market trends. The technical indicators evaluated included Moving Average Convergence Divergence (MACD), Parabolic Stop and Reverse (SAR), Volume-Weighted MACD (VW MACD), and Dual MACD.

As mentioned in Table 2, all the technical indicators exhibited relatively low classification accuracies, with none exceeding 10\%. MACD and SAR both achieved an accuracy of 7.02\%, while VW MACD and Dual MACD had accuracies of 5.26\% and 3.51\%, respectively. These results suggest that, when used in isolation, traditional technical indicators may not provide sufficient predictive power for forecasting stock price movements in our dataset.

\subsection{Combined Indicators Trading Return Analysis}

In this section, we study the added value of sentiment-based strategies as integrated with technical indicators and time-series forecasts. Each model was tested along with an additional filter using sentiment models, technical indicators (MACD, SAR, VM MACD, Dual MACD), and forecasting models (ARIMA, Prophet, ETS), to understand their effect on trading returns. The benchmark represented a vanilla long buy-and-hold strategy that was enacted on the S\&P 500 index.

In Table 3, the highest return observed in this study was achieved by a hybrid strategy that combined sentiment analysis with the VW MACD indicator and time-series forecasts. The return for this combination reached 5.77\%, significantly outperforming the benchmark buy-and-hold strategy, which returned -0.696\%. This demonstrates that sentiment analysis, when combined with robust technical indicators and forecasting models, can provide a substantial advantage in predicting stock price movements and enhancing assets returns.

The negative return relative to the baseline buy-and-hold highlights the need for adaptive models capturing both sentiment and market momentum. Based on our testing, sentiment-based models in isolation will not consistently outperform traditional methods, however, combined with technical indicators significantly enhances the performance.

In Figure 1, the time-series plot of actual vs. predicted daily returns shows that while actual returns (gray line) fluctuate significantly on a day-to-day basis, the predicted returns from ARIMA (green), ETS (red), and Prophet (blue) models remain closer to zero. Both ETS and ARIMA exhibit good alignment with actual daily returns, particularly during periods of high volatility, demonstrating their ability to enhance trading strategies when combined with sentiment signals. The Prophet model, while performing slightly better in some cases(59.65\% accuracy), also provides accurate daily forecasts during periods of lower volatility.

As we can conclude from Figure 2, the dynamic portfolio values of various trading models from May to early August highlight distinct performance patterns over time. The benchmark buy-and-hold strategy, represented by the black line, shows significant volatility and experiences a sharp decline in early August. In contrast, models like "Dual MACD FinBERT Benzinga" (light blue) and "GPT Dow Jones" (green) demonstrate more stable upward trends, effectively avoiding the sharp fluctuations seen in the benchmark. These findings emphasize the potential of combining sentiment analysis with technical indicators to develop adaptive and resilient trading strategies that can mitigate losses and leverage market trends, particularly during periods of heightened volatility.

In Table 3, across five news sources, GPT-2 demonstrates strong performance, particularly on GPT Dow Jones with VW MACD, achieving the highest return of 5.77\%. In Table 4, GPT-2 and FinBERT show varying performance when combined with technical indicators. While FinBERT outperforms GPT-2 in specific cases, such as with MACD and ARIMA for Dow Jones (2.56\% vs. 1.67\%) and with VW MACD ARIMA for WSJ (3.24\% vs. 1.48\%), GPT-2 achieves the highest return of 4.39\% when paired with VW MACD ARIMA for Dow Jones, significantly outperforming FinBERT's 0.52\%. In Table 5, FinBERT's best performance is observed on Benzinga with Dual MACD, reaching 4.64\%. The combination of VW MACD and Dual MACD consistently generates positive returns across different models and news sources, while Prophet and ARIMA models provide more stable but generally lower returns. In summary, while both models show strengths in different scenarios, GPT-2 tends to achieve higher returns, particularly when combined with VW MACD on Dow Jones data, demonstrating the effectiveness of integrating sentiment analysis with technical indicators for optimizing trading performance.

\section{Conclusion}

In this work, we show a way to improve trading strategies in the S\&P 500 market by incorporating sentiment analysis from financial news together with traditional technical indicators and time-series forecasting models. The hybrid strategy triumphed the baseline buy-and-hold by implementing sentiment analysis from GPT-2 and FinBERT combined with other technical indicators such as MACD and SAR, and time-series models like ARIMA and ETS. The top-performing version delivered a 5.77\% return, demonstrating the value that can be had by adding real-time sentiment to stock price forecasting. Financial language nuances were also captured well by sentiment models such as FinBERT, which further increased trading accuracy. This highlights the importance of combining sentiment-based data with technical analysis to further your trading strategy that can be tailored to trade alongside significant movements in price.

\section{Future Work}

There are several directions from which future work could further improve the hybrid strategy. An important one is to use reinforcement learning for trading strategies that adjust themselves on the fly. For instance, reinforcement learning could potentially give the model the ability to learn and improve its trades over time as it becomes better at predicting market performance based on real-time feedback. With reinforcement learning, the different market conditions could be learned and the trading signals could be tweaked accordingly to help optimize risk-reward performance long term.

Moreover, broadening the sentiment sources to listen for real-time data from social media like X or Reddit will give much faster sentiment insights which could be a very good complement to news-based data. Nevertheless, prediction accuracy could be enhanced by applying state-of-the-art deep learning models like LSTM and Transformers to both sentiment analysis and time-series forecasting.

The use of machine learning techniques such as automated feature selection to tune the model, identifying the most impactful sentiment and market features. Generalizing this approach to other asset classes from the financial markets (like commodities or cryptocurrencies) would be a way to check how robust the method is. Live market testing for building a real-time trading system managing the hybrid approach and performance validation that would be free from possible sentiment manipulation as per regulatory and ethical standards.

While our results demonstrate the potential of combining real-time sentiment analysis with technical and time-series models, several limitations warrant discussion. First, we assume zero transaction costs and perfect execution, which likely overstates achievable returns in live markets. Second, our sentiment models treat each headline equally and do not account for article length, author credibility, or source-specific biases. Third, the backtests cover a relatively short period (May–August 2024), so performance in different market regimes (e.g., bear markets or extended rallies) remains untested. Finally, we focus exclusively on the S\&P 500 index; extensions to individual equities or other asset classes may require bespoke model tuning. Addressing these limitations in future work—by incorporating execution costs, refining sentiment weighting schemes, extending the sample period, and broadening asset coverage—would enhance the robustness and generalizability of our findings.

\pagebreak

\section{Figures \& Tables}

\begin{figure}[H]
\includegraphics[width=15cm]{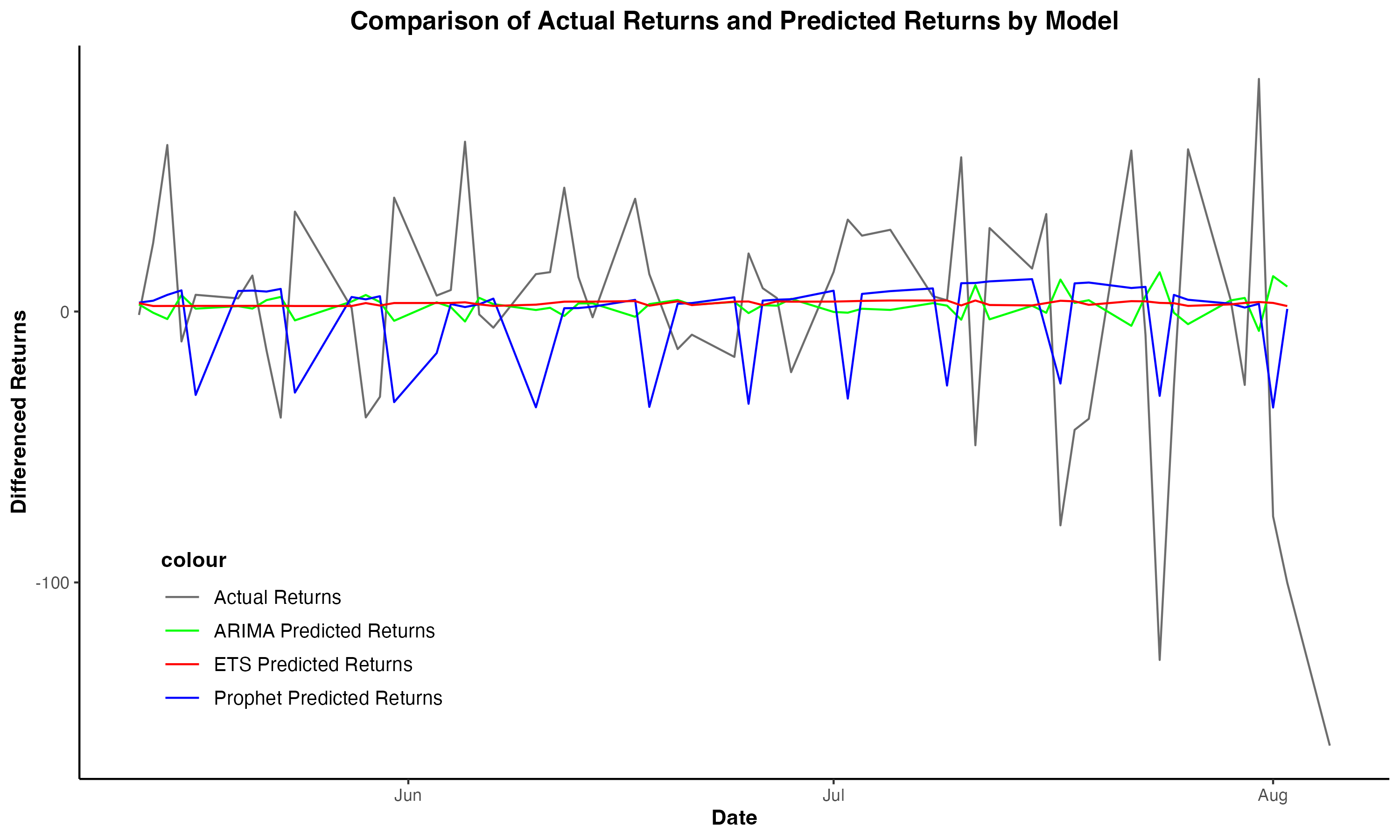}
\centering
\caption{This plot compares the actual returns with predicted returns from ARIMA, ETS, and Prophet models over a period from June to August. The actual returns, represented by the dashed black line, show high variability, while the predicted returns from the models—ARIMA (red), ETS (green), and Prophet (blue)—tend to fluctuate less around zero, with Prophet showing the most noticeable deviations. The graph highlights how each model performs in forecasting, with differences between the actual returns and model predictions visible, particularly during periods of sharp changes in the actual returns.}
\end{figure}

\begin{figure}[H]
\includegraphics[width=15cm]{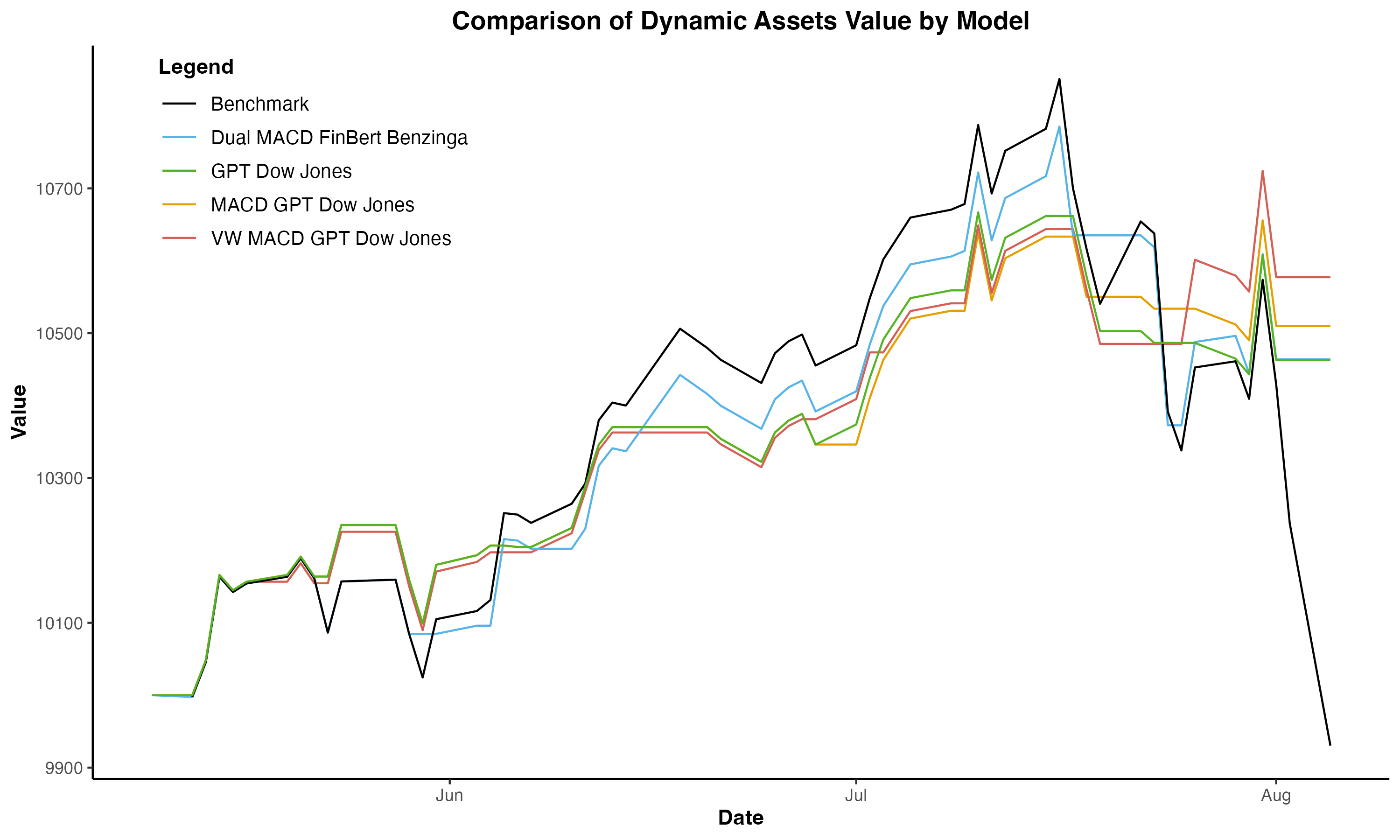}
\centering
\caption{This plot compares the assets values of various trading models from May to early August, showing how each model's value changes over time. The black line, representing the benchmark, is more volatile and experiences a sharp decline in early August. Other models, like the "Dual MACD FinBert Benzinga" (light blue) and "GPT Dow Jones" (green), show more stable upward trends, avoiding the benchmark's sharp fluctuations.}
\end{figure}

\newpage

\begin{center}
    \textbf{News Sentiment Classification Accuracy of FinBERT and GPT-2 Models}
\end{center}

\begin{table}[h!]
\centering
\begin{tabular}{|l|c|c|}
\hline
\textbf{News Source} & \textbf{FinBERT Accuracy (\%)} & \textbf{GPT-2 Accuracy (\%)} \\ \hline
Dow Jones            & 67.69                         & 60.00                        \\ \hline
Benzinga             & \textbf{75.56}                & 64.44                        \\ \hline
Barron               & 50.63                         & 64.56                        \\ \hline
MarketWatch          & 32.22                         & 34.44                        \\ \hline
WSJ                  & 57.14                         & \textbf{65.48}                        \\ \hline
\end{tabular}
\caption{
The table compares the sentiment classification accuracy of the FinBERT and GPT-2 models across five financial news sources. FinBERT outperforms GPT-2 on most sources, achieving the highest accuracy of 75.56\% on Benzinga and 67.69\% on Dow Jones. In contrast, GPT-2 shows better performance on Barron and WSJ, with accuracies of 64.56\% and 65.48\%, respectively. For MarketWatch, both models have lower accuracies, with FinBERT at 32.22\% and GPT-2 at 34.44\%. Overall, the results highlight the strengths of each model in specific news sources, with FinBERT generally showing superior performance.}
\label{tab:sentiment_accuracy}
\end{table}

\noindent
\begin{center}
    \textbf{Time-Series and Traditional Indicator Classification Accuracy}
\end{center}

\begin{table}[h!]
\centering
\begin{tabular}{|l|c|c|}
\hline
\textbf{Type of Indicator} & \textbf{Indicator Accuracy (\%)} \\ \hline
ARIMA            & 29.82                        \\ \hline
Prophet             & 59.65               \\ \hline
ETS               & 59.65                         \\ \hline
MACD          & 7.02                        \\ \hline
SAR                  & 7.02                         \\ \hline
VM MACD                  & 5.26                        \\ \hline
Dual MACD                  & 3.51                         \\ \hline
\end{tabular}
\caption{The table presents the accuracies of various indicators in predicting financial trends. Prophet and ETS models achieve the highest accuracy, both at 59.65\%, followed by ARIMA at 29.82\%. Other indicators, including MACD, SAR, VM MACD, and DUAL MACD, perform significantly worse, with accuracies below 10\%. Notably, none of the accuracies reported in this table exceed those observed in the sentiment classification table, where FinBERT achieved a maximum accuracy of 75.56\% on Benzinga and GPT-2 reached 65.48\% on WSJ. }
\label{tab:indicator_accuracy}
\end{table}

\newpage
\noindent
\begin{center}
    \textbf{Trading Returns of GPT-2 and All Technical Indicators (in \%)}
\end{center}
\begin{table}[h!]
\resizebox{\textwidth}{!}{
\begin{tabular}{|l|c|c|c|c|c|}
\hline
\textbf{MODEL} & \textbf{Dow Jones} & \textbf{Benzinga} & \textbf{Barron} & \textbf{MarketWatch} & \textbf{WSJ} \\
\hline
GPT Dow Jones & 4.63 & 2.40 & -0.67 & 3.83 & 1.13 \\
GPT Benzinga & 2.40 & 2.37 & -0.70 & 2.40 & 2.37 \\
GPT Barron & -0.67 & -0.70 & -0.70 & -0.67 & -0.70 \\
GPT MarketWatch & 3.83 & 2.40 & -0.67 & -0.42 & -2.31 \\
GPT WSJ & 1.13 & 2.37 & -0.70 & -2.31 & -1.06 \\
FinBERT Dow Jones & 0.50 & 2.37 & -0.70 & 1.11 & 0.80 \\
FinBERT Benzinga & 2.60 & 2.37 & -0.70 & 0.70 & 0.35 \\
FinBERT Barron & -1.76 & -0.67 & -0.67 & 3.37 & -2.38 \\
FinBERT MarketWatch & 0.00 & 2.37 & -0.70 & 0.00 & 0.46 \\
FinBERT WSJ & 2.03 & -0.70 & -0.70 & -0.83 & -2.04 \\
ARIMA & 1.67 & 2.37 & 2.37 & 0.75 & 0.20 \\
Prophet & 0.51 & 2.40 & 2.40 & -2.25 & 1.27 \\
ETS & 2.40 & 2.37 & 2.37 & 2.40 & 2.37 \\
MACD & 5.10 & 2.83 & -0.25 & 0.83 & -0.61 \\
SAR & 4.00 & 1.76 & -1.29 & 0.83 & -1.65 \\
VW MACD & \textbf{5.77} & 2.87 & -0.21 & 0.16 & 0.22 \\
Dual MACD & 5.38 & 3.11 & 0.02 & 0.85 & -0.35 \\
\hline
\end{tabular}
}
\caption{This table compares the trading performance of GPT-2 with sentiment indicators from both GPT-2 and FinBERT, as well as various technical indicators (e.g., MACD, SAR, VW MACD, Dual MACD) and time-series models (ARIMA, Prophet, ETS) across five financial news sources (GPT Dow Jones, GPT Benzinga, GPT Barron, GPT MarketWatch, and GPT WSJ).  The highest return, 5.77\%, is achieved by GPT Dow Jones with VW MACD. FinBERT also performs well on GPT Benzinga with Dual MACD, yielding a return of 4.64\%. It shows strong performance on GPT Dow Jones with Dual MACD, achieving a return of 5.38\%, while FinBERT delivers lower returns on GPT Dow Jones and GPT WSJ. Among the models, VW MACD and Dual MACD consistently generate positive returns, while Prophet and ARIMA models provide more stable but generally lower returns. Overall, GPT-2 tends to outperform FinBERT, particularly on GPT Dow Jones and GPT Benzinga.}
\label{tab:trading_performance}
\end{table}

\newpage
\noindent
\begin{center}
    \textbf{Comparison of Trading Returns Between GPT-2 and FinBERT (in \%)}    
\end{center}

\begin{table}[h!]
\centering
\resizebox{\textwidth}{!}{%
\begin{tabular}{|l|c|c|c|c|c|}
\hline
\textbf{Model Indicator} & \textbf{Dow Jones} & \textbf{Benzinga} & \textbf{Barron} & \textbf{MarketWatch} & \textbf{WSJ} \\ \hline
\multirow{2}{*}{MACD ARIMA}    & GPT: 1.67  & GPT: 2.37   & GPT: 2.37  & GPT: 4.01  & GPT: 0.20  \\ 
                               & FinBERT: 2.56  & FinBERT: 2.37  & FinBERT: 2.67  & FinBERT: 2.22   & FinBERT: 1.64 \\ \hline
\multirow{2}{*}{MACD Prophet}  & GPT: 0.24  & GPT: 2.12     & GPT: 2.12  & GPT: -1.62  & GPT: 1.00  \\ 
                               & FinBERT: -0.04  & FinBERT: 0.77  & FinBERT: 2.03  & FinBERT: 0.41  & FinBERT: -1.47 \\ \hline
\multirow{2}{*}{MACD ETS}      & GPT: 2.40 & GPT: 2.37  & GPT: 2.37  & GPT: 2.40   & GPT: 2.37  \\ 
                               & FinBERT: 2.37 & FinBERT: 2.37  & FinBERT: 2.40   & FinBERT: 2.37   & FinBERT: 2.37 \\ \hline
\multirow{2}{*}{SAR ARIMA}     & GPT: 0.53  & GPT: 1.23  & GPT: 1.23 & GPT: 4.01   & GPT: -0.92 \\ 
                               & FinBERT: 1.93  & FinBERT: 1.23 & FinBERT: 1.52 & FinBERT: 2.22   & FinBERT: 1.64 \\ \hline
\multirow{2}{*}{SAR Prophet}   & GPT: 0.51  & GPT: 2.40 & GPT: 2.40 & GPT: -1.62  & GPT: 1.27 \\ 
                               & FinBERT: 0.23   & FinBERT: 1.04 & FinBERT: 2.30  & FinBERT: 0.41  & FinBERT: -0.60 \\ \hline
\multirow{2}{*}{SAR ETS}       & GPT: 2.40  & GPT: 2.37  & GPT: 2.37 & GPT: 2.40   & GPT: 2.37  \\ 
                               & FinBERT: 2.37  & FinBERT: 2.37  & FinBERT: 2.40  & FinBERT: 2.37   & FinBERT: 2.37 \\ \hline
\multirow{2}{*}{VW MACD ARIMA} & GPT: \textbf{4.39}  & GPT: 2.03  & GPT: 2.03  & GPT: 3.59  & GPT: 1.48  \\ 
                               & FinBERT: 0.52  & FinBERT: 1.48  & FinBERT: 2.11 & FinBERT: 0.81    & FinBERT: \textbf{3.24} \\ \hline
\multirow{2}{*}{VW MACD Prophet} & GPT: 3.49 & GPT: 2.30  & GPT: 2.30  & GPT: -0.69   & GPT: -0.05  \\ 
                                 & FinBERT: 0.14 & FinBERT: 1.11  & FinBERT: 2.30   & FinBERT: -3.42   & FinBERT: 2.47 \\ \hline
\multirow{2}{*}{VW MACD ETS}   & GPT: 2.40  & GPT: 2.37  & GPT: 2.37 & GPT: 2.40   & GPT: 2.37  \\ 
                               & FinBERT: 2.37 & FinBERT: 2.37  & FinBERT: 2.40  & FinBERT: 2.37   & FinBERT: 2.37 \\ \hline
\multirow{2}{*}{Dual MACD ARIMA} & GPT: 1.67  & GPT: 2.37 & GPT: 2.37  & GPT: 4.03   & GPT: 0.20 \\ 
                                 & FinBERT: 1.06  & FinBERT: 2.37 & FinBERT: 2.67  & FinBERT: 2.22   & FinBERT: 1.64 \\ \hline
\multirow{2}{*}{Dual MACD Prophet} & GPT: 0.51 & GPT: 2.40 & GPT: 2.40 & GPT: -1.62 & GPT: 1.27 \\ 
                                   & FinBERT: 0.23 & FinBERT: 1.04 & FinBERT: 2.30 & FinBERT: 0.54 & FinBERT: -0.60 \\ \hline
\multirow{2}{*}{Dual MACD ETS} & GPT: 2.40  & GPT: 2.37 & GPT: 2.37  & GPT: 2.40   & GPT: 2.37 \\ 
                               & FinBERT: 2.37  & FinBERT: 2.37 & FinBERT: 2.40  & FinBERT: 2.37  & FinBERT: 2.37 \\ \hline
\end{tabular}%
}
\caption{The table compares the trading performance of GPT-2 and FinBERT models when combined with various technical indicators (e.g., MACD, SAR, VW MACD, Dual MACD) and time-series models (e.g., ARIMA, Prophet, ETS) across different news sources. While FinBERT outperforms GPT-2 in certain cases, such as when paired with MACD and ARIMA for Dow Jones (2.56\% vs. 1.67\%) and with VW MACD ARIMA for WSJ (3.24\% vs. 1.48\%), GPT-2 achieves the highest return overall, 4.39\%, when paired with VW MACD ARIMA for Dow Jones, surpassing FinBERT's 0.52\%. Additionally, GPT-2 outperforms FinBERT with a return of 4.01\% using MACD ARIMA on MarketWatch, compared to FinBERT’s 2.22\%. These results demonstrate that integrating sentiment analysis with technical indicators enhances trading performance, with GPT-2 excelling in specific scenarios, particularly with VW MACD, while FinBERT generally performs better with MACD and ARIMA indicators in some cases.}
\label{tab:trading_performance}
\end{table}

\begin{table}[h!]
\noindent
\begin{center}
    \textbf{Trading Returns of FinBERT and All Technical Indicators (in \%)}
\end{center}
\resizebox{\textwidth}{!}{
\begin{tabular}{|l|c|c|c|c|c|}
\hline
\textbf{MODEL} & \textbf{Dow Jones} & \textbf{Benzinga} & \textbf{Barron} & \textbf{MarketWatch} & \textbf{WSJ} \\
\hline
GPT Dow Jones & 0.50 & 2.60 & -1.76 & 0.00 & 2.03 \\
GPT Benzinga & 2.37 & 2.37 & -0.67 & 2.37 & -0.70 \\
GPT Barron & -0.70 & -0.70 & -0.67 & -0.70 & -0.70 \\
GPT MarketWatch & 1.11 & 0.70 & 3.37 & 0.00 & -0.83 \\
GPT WSJ & 0.80 & 0.35 & -2.38 & 0.46 & -2.04 \\
FinBERT Dow Jones & 2.75 & 0.01 & -0.72 & 1.61 & 2.66 \\
FinBERT Benzinga & 0.01 & 4.07 & -0.67 & 0.85 & 2.90 \\
FinBERT Barron & -0.72 & -0.67 & 0.35 & 0.00 & -1.89 \\
FinBERT MarketWatch & 1.61 & 0.85 & 0.00 & -0.02 & 0.46 \\
FinBERT WSJ & 2.66 & 2.90 & -1.89 & 0.46 & -2.92 \\
ARIMA & 1.06 & 2.37 & 2.67 & 0.46 & 0.99 \\
Prophet & 0.23 & 1.04 & 2.30 & 0.00 & -0.60 \\
ETS & 2.37 & 2.37 & 2.40 & 2.37 & 2.37 \\
MACD & 2.46 & 4.36 & 0.80 & 0.50 & -2.65 \\
SAR & 1.39 & 3.27 & -0.25 & 0.50 & -2.39 \\
VW MACD & 2.95 & 4.58 & 0.21 & -1.14 & -3.08 \\
Dual MACD & 2.75 & \textbf{4.64} & 1.07 & 0.52 & -2.39 \\
\hline
\end{tabular}
}

\caption{This table compares the trading performance of FinBERT with sentiment indicators from both GPT-2 and FinBERT, as well as various technical indicators (e.g., MACD, SAR, VW MACD, Dual MACD) and time-series models (ARIMA, Prophet, ETS) across five financial news sources (GPT Dow Jones, GPT Benzinga, GPT Barron, GPT MarketWatch, and GPT WSJ).  The highest return, 4.64\%, is achieved by FinBERT Benzinga with Dual MACD. It also performs well on FinBERT Dow Jones with Dual MACD, with a return of 2.75\%, and a return of 2.66\% on FinBERT Dow Jones with SAR. GPT-2 shows mixed results, achieving a return of 1.11\% on FinBERT MarketWatch but negative returns on FinBERT Barron and FinBERT WSJ. Among the models, VW MACD and Dual MACD.}
\label{tab:trading_performance}
\end{table}

\clearpage

\bibliography{references}

\end{document}